\documentstyle[12pt]{article}
\makeatletter
\newcount\N  \newcount\M  \newdimen\SIZE  \newdimen\INC
\def\YGBOX#1#2#3{
      \N=#1  \M=1  \INC=#2pt  \advance\INC by .#3pt  
      \vbox{
         \loop\ifnum\M>0
            \M=\N
            \divide\N by 10         
            \multiply\N by 10
            \advance\M by -\N
            \divide\N by 10
            \SIZE=\INC
            \multiply\SIZE by \M    
            \advance\SIZE by .#3pt
             \hrule  width \SIZE  height .#3pt
              \hbox{\loop\ifnum\M>0                      
                        \vrule  height #2pt  width .#3pt 
                        \hskip #2pt
                        \advance\M by -1  \repeat
                        \vrule  width .#3pt }
             \hrule  width \SIZE  height .#3pt
            \vskip -.#3pt
         \repeat } }

\def\young#1{{             
       \mathchoice{\YGBOX{#1}61}{\YGBOX{#1}61}{\YGBOX{#1}41}{\YGBOX{#1}31}}}
\newcount\N  \newcount\M  \newdimen\SIZE  \newdimen\INC
\def\YGBOXC#1#2#3{
      \N=#1  \M=1  \INC=#2pt  \advance\INC by .#3pt  
      \vcenter{\vbox{
         \loop\ifnum\M>0
            \M=\N
            \divide\N by 10         
            \multiply\N by 10
            \advance\M by -\N
            \divide\N by 10
            \SIZE=\INC
            \multiply\SIZE by \M    
            \advance\SIZE by .#3pt
             \hrule  width \SIZE  height .#3pt
              \hbox{\loop\ifnum\M>0                      
                        \vrule  height #2pt  width .#3pt 
                        \hskip #2pt
                        \advance\M by -1  \repeat
                        \vrule  width .#3pt }
             \hrule  width \SIZE  height .#3pt
            \vskip -.#3pt
         \repeat } } }

\newdimen\LENB \newdimen\LENW \newdimen\THI 
\newdimen\LENWH \newdimen\LENTOT \newcount\N 
\def\vbrknlnele#1#2#3{
  \LENB=#1pt \LENW=#2pt \THI=#3pt
  \LENWH=\LENW \divide\LENWH by 2
  \LENTOT=\LENB \advance\LENTOT by \LENW
  \vbox to \LENTOT{
    \vbox to \LENWH{}
    \nointerlineskip
    \vbox to \LENB{\hbox to \THI{\vrule width \THI height \LENB}}
    \nointerlineskip
    \vbox to \LENWH{}
  }}

\def\vbrknln#1{
  \N=#1
  \vcenter{
    \vbox{
      \loop\ifnum\N>0
        \vbox to 4pt{\vbrknlnele{2}{2}{0.1}}
        \nointerlineskip
        \advance\N by -1
      \repeat
  }}}

\def\hbrknlnele#1#2#3{
  \LENB=#1pt \LENW=#2pt \THI=#3pt
  \LENTOT=\LENB \advance\LENTOT by \LENW
  \vcenter{
    \vbox to \THI{
      \hbox to \LENTOT{
        \hfil
        \vrule width \LENB height \THI
        \hfil}
  }}}

\def\hblele{\hbrknlnele{2}{2.2}{0.1}}
\def\hblfil{\cleaders\hbox{$ \m@th \mkern1mu \hblele \mkern1mu
$}\hfill} 
\makeatother
\newtheorem{th}{{\sc Theorem}}[section]
\newtheorem{prop}[th]{{\sc Proposition}}
\newtheorem{df}[th]{{\sc Definition}}
\newtheorem{lem}[th]{{\sc Lemma}}
\newtheorem{rem}[th]{{\sc Remark}}

\renewcommand{\thesection}{\Roman{section}.}
\begin{document}

\begin{titlepage}
\begin{center}
\begin{Large}
{\bf Determinant Structure of the Rational Solutions}\\
{\bf for the Painlev\'e II Equation}\\
\end{Large}
\vspace{30pt}
\begin{normalsize}
{\sc Kenji Kajiwara}\\
{\it Department of Electrical Engineering,
Doshisha University,}\\
{\it Tanabe, Kyoto 610-03, Japan}\\
and \\
{\sc Yasuhiro Ohta}\\
{\it Department of Applied Mathematics, Faculty of Engineering,}\\
{\it Hiroshima University, }\\
{\it 1-4-1 Kagamiyama, Higashi-Hiroshima 739, Japan}\\
\end{normalsize}
\end{center}
\vspace{30pt}
\begin{abstract}
Two types of determinant representations of the rational solutions
for the Painlev\'e II equation 
are discussed by using the bilinear formalism.
One of them is a representation by the Devisme polynomials,
and another one is a Hankel determinant representation.
They are derived from the determinant solutions
of the KP hierarchy and Toda lattice, respectively.
\end{abstract}
\end{titlepage}

\addtolength{\baselineskip}{.5\baselineskip}

\section{Introduction}
The six Painlev\'e transcendents are now regarded as the nonlinear
version of the special functions and hence
the Painlev\'e equations are the most fundamental
integrable systems in some sense.
It is known that the Painlev\'e transcendents
cannot be expressed by
the solutions of linear equations, except for two classes of
solutions,
namely, special function solutions and rational solutions.
The Painlev\'e II equation(P$_{\rm II}$)
\begin{equation}
\frac{d^2}{dz^2}v = 2v^3 - 4zv + 4\alpha,
\label{PII}
\end{equation}
is the simplest equation that admits such solutions among the
Painlev\'e equations.
In fact, it is known that it admits one parameter family of Airy function solutions
for $\alpha$ being half odd integers, and only one rational solutions
for each integer $\alpha$\cite{Okamoto} and it has no other classical solution\cite{Umemura}. \par
It is well known that the Painlev\'e equations can be derived
from the similarity reduction of various soliton equations\cite{Ablowitz}.
In particular, P$_{\rm II}$ can be reduced from the modified KdV equation.
A systematic study of
the rational solutions was done by Airault\cite{Airault}, who
constructed the B\"acklund transformation of P$_{\rm II}$ from
the similarity reduction of the modified KdV equation.
On the other hand, Okamoto revealed that the B\"acklund transformations
of Painlev\'e equations are given by the Toda lattice equation.
For the KP and Toda lattice hierarchies, the solutions
are described by the Wronski determinants.
The Painlev\'e equations are deeply connected with the KP and Toda,
therefore a question naturally arising is
what is the structure of the solutions of Painlev\'e equations.
Actually for the special function type solutions of Painlev\'e equations,
it is known that they are
expressed by Wronskians whose entries are given by special functions.
Such Wronskians are called the $\tau$ function. Here, we note that $\tau$ functions are
originally defined for arbitrary values of parameters
through the Hamiltonians of the Painlev\'e equations\cite{Okamoto},\cite{JMU}.
This Wronskian structure of the solutions is quite similar to
that for the soliton
equations. Hence we expect that the rational solutions also have
such a structure.
Many studies have been done for the rational solutions, but curiously,
it seems that the determinant structure of solutions itself
has not been well discussed. This situation motivates us for studying the relationship
of the solutions of the Painlev\'e equations and integrable PDE.

In this article, we present the determinant representations for 
the rational solutions of P$_{\rm II}$
and clarify how those solutions are reduced from the $\tau$ functions
of the KP hierarchy and Toda lattice.
We present two types of determinant representations. One is directly derived
from the Schur polynomials, namely the algebraic solutions for the
KP hierarchy,
by applying a reduction procedure. Entries of the determinant
are expressed by the Devisme polynomials\cite{Devisme},\cite{Erd}.
This reduction exactly corresponds to the derivation of P$_{\rm II}$
from the modified KdV equation.
The bilinear form for P$_{\rm II}$ is nothing but the bilinear
first B\"acklund transformations of the KP hierarchy.
Another one is a Hankel determinant representation
which is derived from the Hankel determinant solution
of the B-type Toda lattice equation\cite{UandT}.
In this case, the Toda lattice is corresponding to
the B\"acklund transformation ladder of the solutions of P$_{\rm II}$.

In section II, the bilinearization of P$_{\rm II}$ is presented.
We give a brief review of the algebraic solutions
for KP and KdV hierarchies in section III.
In section IV, we give the derivation of the 
rational solutions for P$_{\rm II}$ from the Schur polynomials.
In section V, we briefly summarize the determinant solution
for the B-type Toda lattice equation.
The Hankel determinant representation of the rational solutions
is presented in section VI. Section VII is devoted
to concluding remarks.

\section{Bilinear form for P$_{\rm II}$}
By using the dependent variable transformation,
\begin{equation}
v=\frac{d}{dz}\log\frac{g}{f},
\label{ovartr}
\end{equation}
eq.(\ref{PII}) is decomposed into
the following bilinear equations\cite{JMU},\cite{Hietarinta},
\begin{eqnarray}
&&\left( D_z^2 - \lambda \right) g\cdot f = 0,
\label{obl1l}\\
&&\left( D_z^3 + (4z-3\lambda)D_z - 4\alpha \right) g\cdot f = 0,
\label{obl2l}
\end{eqnarray}
where $D_z^n$ is the Hirota bilinear differential operator 
and $\lambda$ is an arbitrary function of $z$.
Dividing eqs.(\ref{obl1l}) and (\ref{obl2l}) by $gf$, 
we obtain
\begin{eqnarray}
&&s + v^2 = \lambda,
\nonumber\\
&&v_{zz} + 3sv + v^3 + (4z-3\lambda)v - 4\alpha = 0,
\nonumber
\end{eqnarray}
where $s=(\log gf)_{zz}$.  
Eliminating $s$ from above equations, 
we get P$_{\rm II}$ (\ref{PII}),
therefore eqs.(\ref{obl1l}) and (\ref{obl2l}) actually give 
the bilinear form for P$_{\rm II}$.

Using the gauge transformation, 
we can take $\lambda$ as we like.  
In the case of the rational solutions of P$_{\rm II}$, 
taking $\lambda$ to be $0$ is convenient as is shown in the later.  
On the other hand, 
for the Airy function type solutions of P$_{\rm II}$, 
$\lambda$ is taken to be $2z$\cite{Okamoto}.  
If we fix $\lambda$ to be equal $0$, 
the bilinear equations for P$_{\rm II}$ are 
\begin{eqnarray}
&&D_z^2 g\cdot f = 0,
\label{obl1}\\
&&\left( D_z^3 + 4zD_z - 4\alpha \right) g\cdot f = 0.
\label{obl2}
\end{eqnarray}
In this gauge, 
these equations allow polynomial solutions for $f$ and $g$ 
which give the rational solutions $v$ for P$_{\rm II}$ (\ref{PII}) 
through the variable transformation (\ref{ovartr}).  
In the following sections, 
we will show how the rational solutions are constructed from 
the $\tau$ functions of KP hierarchy and Toda lattice equation.  

\section{Algebraic Solutions for KP and KdV Hierarchies}
We first give a brief review on the algebraic solutions of KP and
KdV hierarchies\cite{JM}. 
\renewcommand{\thesection}{\arabic{section}}
\begin{df}
Let $p_j(y)$, $j=0,1,2,\cdots$,
be polynomials in $y=(y_1,y_2,y_3,\cdots)$ defined by
\begin{equation}
\sum_{k=0}^\infty p_k(y)\lambda^k={\rm exp}\sum_{n=1}^\infty
y_n\lambda^n\ ,\quad {\rm and }\ \ p_k(y)=0,\ {\rm for}\ k<0\ .\label{elementary_Schur}
\end{equation}
Then a set of infinitely many bilinear equations for
$\tau(x)=\tau(x_1,x_2,x_3,\cdots) $ generated by
\begin{equation}
\left( \sum_{j=0}^{\infty}p_j(-2y)p_{j+1}(\tilde D)\exp \left(
\sum_{n=1}^\infty y_nD_{x_n}\right)\right)\tau\cdot\tau=0\ ,
\end{equation}
where
\[ \tilde D = (D_{x_1},\frac{1}{2}D_{x_2}, \frac{1}{3}D_{x_3},\cdots)\ ,\]
is called the KP hierarchy and $\tau$ is called the $\tau$ function.
\end{df}
The simplest bilinear equation included in this hierarchy is
\begin{equation}
\left( D_{x_1}^4 - 4D_{x_1}D_{x_3}+3D_{x_2}^2\right)\tau\cdot\tau=0\ ,
\end{equation}
which yields the KP equation in nonlinear form,
\begin{equation}
\left (-4 u_{x_3} + 6uu_{x_1}+u_{x_1x_1x_1}\right)_{x_1}+3u_{x_2}=0\ ,\label{KP}
\end{equation}
by the dependent variable transformation,
\begin{equation}
u=2\left(\log \tau\right)_{x_1x_1}\ .\label{dep_KP}
\end{equation}
\begin{prop}
The following Wronskian,
\begin{equation}
\tau_{N,{\rm KP}}=\left|\matrix{
\partial_{x_1}^{N-1}f_1 & \cdots & \partial_{x_1}f_1 & f_1\cr
\partial_{x_1}^{N-1}f_2 & \cdots & \partial_{x_1}f_2 & f_2\cr
\vdots                  & \cdots & \vdots            & \vdots\cr
\partial_{x_1}^{N-1}f_N & \cdots & \partial_{x_1}f_N & f_N\cr}\right|, 
\label{tau_KP}
\end{equation}
solves the KP hierarchy, where
$f_k$, $k=1,2,\cdots,N$ are arbitrary functions in infinitely many
independent
variables $x=(x_1, x_2, \cdots)$ satisfying
\begin{equation}
\partial_{x_n}f_k=\partial_{x_1}^nf_k,\quad k=1,2,\cdots,N,\
\ n=1,2,\cdots\ .\label{dispersion}
\end{equation}
\end{prop}

The crucial point is that all the bilinear equations in the KP
hierarchy for the $\tau$ function (\ref{tau_KP}) 
are reduced to the identities of determinant which are called
the Pl\"ucker relations. 
\begin{df}
A set of infinitely many bilinear equations in
$\tau(x)$ and $\tau^\prime(x)$ generated by
\begin{equation}
\left( \sum_{j=0}^{\infty}p_j(-2y)p_{j+2}(\tilde D)\exp \left(
\sum_{n=1}^\infty y_nD_{x_n}\right)\right)\tau\cdot\tau^\prime=0\ ,
\end{equation}
is called the first modified KP hierarchy.
\end{df}
In particular, $\tau=\tau_{N+1,{\rm KP}}$, $\tau^\prime=\tau_{N,{\rm KP}}$ solves the
first modified KP hierarchy. Hence, this is regarded as the hierarchy of
the first B\"acklund transformations.
Moreover, the bilinear equations in this hierarchy are regarded as the
identities of $(N+1)\times(N+1)$ determinant and
$N\times N$ determinant, which are also the Pl\"ucker relations.  
First two equations of
this hierarchy are given by
\begin{equation}
\left(D_{x_1}^2 - D_{x_2}\right)\tau_{N+1,{\rm KP}}\cdot\tau_{N,{\rm KP}}=0\ ,
\label{mKP1}
\end{equation}
\begin{equation}
\left(D_{x_1}^3 - 4D_{x_3} + 3D_{x_1}D_{x_2}\right)\tau_{N+1,{\rm KP}}
\cdot\tau_{N,{\rm KP}}=0 \ .\label{mKP2}
\end{equation}
In the following, 
we show that these two equations reduce to the bilinear form 
for P$_{\rm II}$ (\ref{obl1}) and (\ref{obl2}) on the conditions of 
reduction for the algebraic solutions.  

Now we discuss the algebraic solutions for the KP hierarchy.
We can easily verify that the polynomials $p_k(x)$ defined by
eq.(\ref{elementary_Schur}) satisfy
\begin{equation}
\partial_{x_n}p_k(x)=p_{k-n}(x),\label{schur}
\end{equation}
and hence eq.(\ref{dispersion}). 
Taking $f_k$ in the $\tau$ function (\ref{tau_KP}) as $p_{i_k+N-k}(x)$, 
we have,
\renewcommand{\thesection}{\arabic{section}}
\begin{prop}
Let $Y=(i_1,i_2,\cdots,i_N)$, where $i_1\ge i_2\ge \cdots \ge i_N\ge
0$ are integers,
be a Young diagram. Then 
\begin{equation}
\tau_{Y,{\rm KP}} = \left|\matrix{
p_{i_1}(x) & p_{i_1+1}(x) & \cdots & p_{i_1+N-1}(x)\cr
p_{i_2-1}(x) & p_{i_2}(x) & \cdots & p_{i_2+N-2}(x)\cr
\vdots  & \vdots    & \ddots & \vdots \cr
p_{i_N-N+1}(x) & p_{i_N-N+2}(x) & \cdots & p_{i_N}(x)\cr}\right|\ , 
\label{KPtau}
\end{equation}
gives the algebraic solution for the KP hierarchy.
\end{prop}
The polynomial $\tau_{Y,{\rm KP}}$ is called the Schur polynomial attached
to the Young diagram $Y$.
We note that if we define the weight of $x_n$ as $n$,
then $p_k(x)$ is a polynomial with homogeneous weight $k$ and 
$\tau_{Y,{\rm KP}}$ is also homogeneous with the weight 
$|Y|=i_1+i_2+\cdots+i_N$. This $\tau$ function gives the rational
solution of eq.(\ref{KP}) by the dependent variable transformation (\ref{dep_KP}).

Let us apply the reduction to the KdV hierarchy. This is achieved
by dropping the dependence of $x_2, x_4,\cdots,$ in the $\tau$ functions 
of KP hierarchy.
In order to realize this condition, 
it is sufficient to choose $Y$ as $(N,N-1,\cdots,1)$
in the algebraic solution for the KP hierarchy (\ref{KPtau}).  
\begin{prop}\label{KdV}
\begin{equation}
\tau_{N,{\rm KdV}} = \left|\matrix{
p_{N}(x) & p_{N+1}(x) & \cdots & p_{2N-1}(x)\cr
p_{N-2}(x) & p_{N-1}(x) & \cdots & p_{2N-3}(x)\cr
\vdots  & \vdots    & \ddots & \vdots \cr
p_{-N+2}(x) & p_{-N+3}(x) & \cdots & p_{1}(x)\cr}\right|\ . \label{KdVtau}
\end{equation}
gives the algebraic solution of the KdV hierarchy.
\end{prop}
Proposition \ref{KdV} can be easily verified noticing that
\begin{equation}
\frac{\partial \tau_{N,{\rm KdV}}}{\partial x_{2j}}=0\ ,\quad
j=1,2,3\cdots\ ,
\label{o2red}
\end{equation}
which directly follows from eq.(\ref{schur}).
{}From eqs.(\ref{mKP1}), (\ref{mKP2}) and (\ref{o2red}), 
it is clear that the $\tau$ function (\ref{KdVtau}) satisfies the
following bilinear equations,
\begin{equation}
D_{x_1}^2\tau_{N+1,{\rm KdV}}\cdot\tau_{N,{\rm KdV}}=0\ ,\label{2-mKP1}
\end{equation}
\begin{equation}
\left(D_{x_1}^3 - 4D_{x_3}\right)\tau_{N+1,{\rm KdV}}
\cdot\tau_{N,{\rm KdV}}=0 \ .\label{2-mKP2}
\end{equation}
The modified KdV equation,
\begin{equation}
v_{x_3}+\frac{3}{2}v^2v_{x_1}-\frac{1}{4}v_{x_1x_1x_1}=0,
\end{equation}
is obtained from eqs.(\ref{2-mKP1}) and (\ref{2-mKP2}) by
the dependent variable transformation,
\begin{equation}
v=\frac{\partial}{\partial x_1}\log\frac{\tau_{N+1,{\rm KdV}}}{\tau_{N,{\rm KdV}}}\ .
\end{equation}
\renewcommand{\thesection}{\Roman{section}.}
\section{Rational Solutions for P$_{\rm II}$: Devisme Polynomial Representation}
Now we give a determinant representation for the rational solutions of
P$_{\rm II}$. We first give the definition of the Devisme polynomials\cite{Devisme},\cite{Erd}.
\renewcommand{\thesection}{\arabic{section}}
\begin{df}
The Devisme polynomials 
$q_k(x_1,x_2,\cdots, x_m)$, $k=0,1,2\cdots$, are polynomials in
$x_1,\cdots, x_m$ defined by
\begin{equation}
\sum_{k=0}^\infty q_k(x_1,x_2,\cdots, x_m)\lambda^k=
{\rm exp}\left( x_1\lambda +
x_2\lambda^2+\cdots+x_m\lambda^m+\frac{1}{m+1}\lambda^{m+1}\right)
\ .
\end{equation}
\end{df}
Then one of our main results is stated as follows.
\renewcommand{\thesection}{\arabic{section}}
\begin{th}\label{main}
Let $q_k(z,t)$, $k=0,1,2,\cdots$, be the Devisme polynomials
and $\tau_N$ be an $N\times N$ determinant defined by
\begin{equation}
\tau_N=\left|\matrix{
q_{N}(z,t) & q_{N+1}(z,t) & \cdots & q_{2N-1}(z,t)\cr
q_{N-2}(z,t) & q_{N-1}(z,t) & \cdots & q_{2N-3}(z,t)\cr
\vdots  & \vdots    & \ddots & \vdots \cr
q_{-N+2}(z,t) & q_{-N+3}(z,t) & \cdots & q_{1}(z,t)\cr}\right|\ ,
\ q_k(z,t)=0\ {\rm for}\ k<0\ . \label{P2tau}
\end{equation}
Then 
\begin{equation}
v=\frac{d}{dz}\log\frac{\tau_{N+1}}{\tau_{N}},\label{var}
\end{equation}
gives a rational solution for  P$_{II}$ (\ref{PII}) with $\alpha=N+1$.
\end{th}
\begin{rem}
\begin{enumerate}
\item 
The $\tau$ function (\ref{P2tau})
is derived only by putting 
\begin{equation}
x_1=z,\quad x_2=t,\quad x_3=\frac{1}{3},\quad x_4=x_5=\cdots=0\ ,
\end{equation}
in eq.(\ref{KdVtau}). Namely, the rational solutions of P$_{\rm II}$ are given in terms of
the special case of the Schur polynomials.
\item The $\tau$ function (\ref{P2tau}) itself
does not depend on $t$, but we have left $t$ dependence
in the entries in order to relate the solutions with the Devisme
polynomials.
\end{enumerate}
\end{rem}
Theorem \ref{main} is a direct consequence of the following proposition.
\begin{prop}
The $\tau$ function $\tau_N$ (\ref{P2tau}) satisfies
the bilinear equations (\ref{obl1}) and (\ref{obl2}) with
$\alpha=N+1$.
\end{prop}
{\it Proof.} 
Putting $x_5=x_7=\cdots=0$ in the rational solutions of KdV hierarchy 
(\ref{KdVtau}), it is readily seen
that $\tau_{N,{\rm KdV}}$
is a homogeneous weight polynomial in $x_1$ and $x_3$ with weight
$\frac{N(N+1)}{2}$. Hence, if we put 
\begin{equation}
f_N=\frac{1}{x_3^{N(N+1)/6}}\tau_{N,{\rm KdV}}\ ,
\end{equation}
then $f_N$ depends only on $t=\frac{\displaystyle x_1}{\displaystyle x_3^{1/3}}$.
Thus we have
\begin{equation}
\partial_{x_3}f_N = \frac{\partial t}{\partial {x_3}}\frac{d}{dt} f_N\ 
, \quad\partial_{x_1}f_N = \frac{\partial t}{\partial {x_1}}\frac{d}{dt} f_N\ ,
\end{equation}
which yield
\begin{equation}
\partial_{x_3}\tau_{N,{\rm KdV}}
=\frac{1}{3x_3}\left(\frac{N(N+1)}{2}\tau_{N,{\rm KdV}}
-x_1\partial_{x_1}\tau_{N,{\rm KdV}}\right). \label{x3}
\end{equation}
Substituting eq.(\ref{x3}) into eq.(\ref{2-mKP2}), we get
\begin{equation}
\left(D_{x_1}^3 + \frac{4}{3x_3}x_1D_{x_1} - \frac{4}{3x_3}(N+1)\right)
\tau_{N+1,{\rm KdV}}
\cdot\tau_{N,{\rm KdV}}=0 \ .
\label{oredbl}
\end{equation}
Moreover, by putting $z=x_1$ and $x_3=\displaystyle{\frac{1}{3}}$,
$\tau_{N,{\rm KdV}}$ reduces to $\tau_N$ in (\ref{P2tau})
and eqs.(\ref{obl1}) and
(\ref{obl2}) with $f=\tau_N$, $g=\tau_{N+1}$ and $\alpha=N+1$ are
obtained
from eqs.(\ref{2-mKP1}) and (\ref{oredbl}).\hfill\lower5pt\hbox{$\Box$}
\renewcommand{\thesection}{\Roman{section}.}
\section{Hankel Determinant Solution for Toda Lattice}

Let us consider the Toda lattice equation,
\begin{equation}
\frac{d^2u_N}{dz^2}={\rm e}^{u_{N-1}-u_N} - {\rm e}^{u_{N}-u_{N+1}}\ ,\label{BToda}
\end{equation}
with the symmetric lattice condition,
\begin{equation}
u_N = u_{-N-1}.
\end{equation}
It is easy to see that eq. (\ref{BToda}) is bilinearized through
the dependent variable transformation,
\begin{equation}
u_N=\log \frac{\tau_{N-1}}{\tau_N}\ ,
\end{equation}
from which we get
\begin{equation}
D_z^2 f_N\cdot f_N = 2( f_{N+1}f_{N-1} - f_Nf_N ),\quad f_N = f_{-N-1}.
\label{oBTodaf}
\end{equation}
Here we call this type of symmetric lattice as B-type Toda lattice
because it concerns the BKP hierarchy\cite{UandT},\cite{JM}.
Using the gauge freedom,
we can translate the above B-type Toda lattice equation
in the following form,
\begin{equation}
(D_z^2 + 2a_0) \sigma_N\cdot\sigma_N
= 2\sigma_{N+1}\sigma_{N-1},
\label{oblToda}
\end{equation}
where $\sigma_N=f_N/f_0$ and $a_0=f_1/f_0$.
It is clear that $\sigma_N$ satisfies
\begin{equation}
\sigma_N = \sigma_{-N-1},
\label{oBToda}
\end{equation}
\begin{equation}
\sigma_0 = 1,\qquad
\sigma_1 = a_0.
\label{obcToda}
\end{equation}
It is possible to express the general solution of
eqs.(\ref{oblToda})-(\ref{obcToda}) in determinant form\cite{Ohta}.
\renewcommand{\thesection}{\arabic{section}}
\begin{prop}
The general solution for the equations
(\ref{oblToda})-(\ref{obcToda}) for an arbitrary $a_0$ is given in the
Hankel determinant form
\begin{equation}
\sigma_N = \left|\matrix{
a_0    &a_1    &\cdots &a_{N-1}\cr
a_1    &a_2    &\cdots &a_N    \cr
\vdots &\vdots &\ddots &\vdots \cr
a_{N-1}&a_N    &\cdots &a_{2N-2}}\right|,
\qquad N\ge0
\label{Hankel}
\end{equation}
where $a_n$, $n=1,2,3,\cdots$ are recursively defined by
\begin{equation}
a_{n+1} = \frac{da_n}{dz} + \sum_{k=0}^{n-1}a_ka_{n-k-1},
\qquad n\geq 0.
\label{orecursion}
\end{equation}
\end{prop}
This contains one arbitrary function $a_0$, hence
it gives the general solution for the B-type Toda lattice equation.

\renewcommand{\thesection}{\Roman{section}.}
\section{Rational Solutions for P$_{\rm II}$: Hankel Determinant Representation}
The rational solutions for P$_{\rm II}$ are derived only by putting
\begin{equation}
a_0 = z,
\label{oini}
\end{equation}
in the above $\sigma_N$.
\renewcommand{\thesection}{\arabic{section}}
\begin{th}\label{main2}
Let $a_n$, $n=0,1,2,\cdots$, be polynomials defined by
\begin{equation}
a_{n+1} = \frac{da_n}{dz} + \sum_{k=0}^{n-1}a_ka_{n-k-1},
\qquad n\geq 0,\quad a_0=z,
\end{equation}
and let $\sigma_N$ be an $N\times N$ determinant given by eq.(\ref{Hankel}).
Then
\begin{equation}
v=\frac{d}{dz}\log\frac{\sigma_{N+1}}{\sigma_N},
\label{dep}
\end{equation}
gives a rational solution for P$_{II}$ (\ref{PII}) with $\alpha=N+1$.
\end{th}

Similar to the previous section, theorem \ref{main2} is a direct consequence of
the following proposition.

\begin{prop}\label{prop2}
$f=\sigma_N$ and $g=\sigma_{N+1}$  satisfies
the bilinear equations (\ref{obl1}) and (\ref{obl2}) with
$\alpha=N+1$.
\end{prop}

To prove proposition \ref{prop2}, let us first introduce the notation $\sigma_{NY}$:

\begin{df}
Let $Y=(i_1,i_2,\cdots,i_h)$ be a Young diagram.
Then we define an $N\times N$ determinant $\sigma_{NY}$ by
\begin{equation}
\sigma_{NY} = \left|\matrix{
a_0    &a_1    &\cdots &a_{N-h-1} &a_{N-h+i_h}
 &\cdots &a_{N-2+i_2} &a_{N-1+i_1}\cr
a_1    &a_2    &\cdots &a_{N-h}   &a_{N-h+1+i_h}
 &\cdots &a_{N-1+i_2} &a_{N+i_1}  \cr
\vdots &\vdots &\cdots &\vdots    &\vdots
 &\cdots &\vdots      &\vdots     \cr
a_{N-1}&a_N    &\cdots &a_{2N-h-2}&a_{2N-h-1+i_h}
 &\cdots &a_{2N-3+i_2}&a_{2N-2+i_1}}\right|.
\end{equation}
\end{df}

We first construct the shift operators which are differential operators
generating $\sigma_{NY}$ from $\sigma_N$.
If entries of the determinant satisfy simple equations like
(\ref{dispersion}), then construction of the shift operators is
straightforward.
But when we have to work on more complicated
relations among the entries like (\ref{orecursion}),
it is useful to apply the technique
developed in \cite{OandN}.

We can prove the following lemma.
\begin{lem}\label{le1}
\begin{equation}
\sigma_N{}_{\young{1}}=\frac{d}{dz}\sigma_N.
\label{shift1}
\end{equation}
\end{lem}
{\it Proof.} Notice that $\sigma_{N\young{1}}$
is expressed by
\begin{equation}
\sigma_{N\young{1}} = \pmatrix{
a_1 & a_2 &\cdots & a_N\cr
a_2 & a_3 &\cdots & a_{N+1}\cr
\vdots &\vdots &\ddots &\vdots\cr
a_N & a_{N+1} &\cdots & a_{2N-1}\cr}\cdot
\pmatrix{
\Delta_{11} & \Delta_{12} &\cdots &\Delta_{1N}\cr
\Delta_{21} & \Delta_{22} &\cdots &\Delta_{2N}\cr
\vdots      & \vdots      &\ddots &\vdots\cr
\Delta_{N1} & \Delta_{N2} &\cdots &\Delta_{NN}\cr}\ ,\label{eq1}
\end{equation}
where $\Delta_{ij}$ is the $(i,j)-$cofactor of $\sigma_N$ and
$A\cdot B$ denotes a standard scalar product for $N\times N$
matrices $A=(a_{ij})$ and $B=(b_{ij})$ which is defined as
\begin{equation}
A\cdot B=\sum_{i,j=1}^N a_{ij}b_{ij}={\rm trace} A^tB.
\end{equation}
The first matrix of
(\ref{eq1}) is rewritten by using the recursion relation
(\ref{orecursion}) as 
\begin{eqnarray}
&{}&
\pmatrix{
\partial_z a_0 &\partial_z a_1 &\cdots &\partial_za_{N-1}\cr
\partial_z a_1 &\partial_z a_2 &\cdots &\partial_za_N    \cr
\vdots         &\vdots         &\ddots &\vdots           \cr
\partial_z a_{N-1} &\partial_z a_N &\cdots &\partial_za_{2N-2}}
\nonumber\\
&+&
\pmatrix{
0      & a_0^2         &\cdots &{\displaystyle\sum_{k=0}^{N-2}}a_ka_{N-k-2}\cr
a_0^2  & a_0a_1+a_1a_0 &\cdots &{\displaystyle\sum_{k=0}^{N-1}}a_ka_{N-k-1}\cr
\vdots & \vdots        &\ddots &\vdots                                     \cr
{\displaystyle\sum_{k=0}^{N-2}}a_ka_{N-k-2}
&{\displaystyle\sum_{k=0}^{N-1}}a_ka_{N-k-1} &\cdots
&{\displaystyle\sum_{k=0}^{2N-3}}a_ka_{2N-k-3}}.
\label{eq2}
\end{eqnarray}
The above second term is separated as
\begin{eqnarray}
&{}&
\pmatrix{
0      &a_0^2  &\cdots &{\displaystyle\sum_{k=0}^{N-2}}a_ka_{N-k-2}\cr
0      &a_0a_1 &\cdots &{\displaystyle\sum_{k=0}^{N-2}}a_ka_{N-k-1}\cr
\vdots &\vdots &\vdots &\vdots                      \cr
0      &a_0a_{N-1} &\cdots &{\displaystyle\sum_{k=0}^{N-2}}a_ka_{2N-k-3}}
\nonumber\\
&+&
\pmatrix{
0      &0      &\cdots &0         \cr
a_0^2  &a_1a_0 &\cdots &a_{N-1}a_0\cr
\vdots &\vdots &\cdots &\vdots    \cr
{\displaystyle\sum_{k=0}^{N-2}}a_ka_{N-k-2}
&{\displaystyle\sum_{k=1}^{N-1}}a_ka_{N-k-1} &\cdots
&{\displaystyle\sum_{k=N-1}^{2N-3}}a_ka_{2N-k-3}}.
\label{eq3}
\end{eqnarray}
Each of these terms gives zero contribution in (\ref{eq1}).
Hence we have proved lemma \ref{le1}.\hfill\lower5pt\hbox{$\Box$}

Next we have:
\begin{lem}\label{le2}
\begin{equation}
\sigma_N{}_{\young{2}}+\sigma_N{}_{\young{11}}
=\left(\frac{d^2}{dz^2}+z\right)\sigma_N\ ,
\label{shift2}
\end{equation}
\begin{equation}
\sigma_N{}_{\young{2}}-\sigma_N{}_{\young{11}}
=(2N-1)z\sigma_N\ .
\label{shift3}
\end{equation}
\end{lem}
{\it Proof.} We consider
\begin{eqnarray}
&{}&\sigma_{N}{}_{\young{2}}+\sigma_{N}{}_{\young{11}}
\nonumber\\
&{}&\qquad
=\pmatrix{
a_1 & a_2 &\cdots & a_{N-1}& a_{N+1}\cr
a_2 & a_3 &\cdots &a_N& a_{N+2}\cr
\vdots &\vdots &\vdots &\vdots&\vdots\cr
a_{N} & a_{N+1} &\cdots &a_{2N-2}& a_{2N}\cr}\cdot
\pmatrix{
\Delta_{\young{1}}{}_{11} & \Delta_{\young{1}}{}_{12} &\cdots &\Delta_{\young{1}}{}_{1N}\cr
\Delta_{\young{1}}{}_{21} & \Delta_{\young{1}}{}_{22} &\cdots &\Delta_{\young{1}}{}_{2N}\cr
\vdots      & \vdots      &\ddots &\vdots\cr
\Delta_{\young{1}}{}_{N1} & \Delta_{\young{1}}{}_{N2} &\cdots
&\Delta_{\young{1}}{}_{NN}\cr}
\ ,
\end{eqnarray}
where $\Delta_{\young{1}}{}_{ij}$ is $(i,j)$ cofactor of
$\sigma_N{}_{\young{1}}$.
The first matrix in the right-hand side is equal to
\begin{eqnarray}
&{}&\pmatrix{
\partial_z a_0 &\partial_z a_1 &\cdots &\partial_za_{N-2} &\partial_za_N    \cr
\partial_z a_1 &\partial_z a_2 &\cdots &\partial_za_{N-1} &\partial_za_{N+1}\cr
\vdots         &\vdots         &\vdots &\vdots            &\vdots           \cr
\partial_z a_{N-1} &\partial_z a_N &\cdots &\partial_za_{2N-3}
&\partial_za_{2N-1}}
\nonumber\\
&+&
\pmatrix{
0      &a_0^2      &\cdots
&{\displaystyle\sum_{k=0}^{N-3}}a_ka_{N-k-3}
&{\displaystyle\sum_{k=0}^{N-1}}a_ka_{N-k-1}\cr
0      &a_0a_1     &\cdots
&{\displaystyle\sum_{k=0}^{N-3}}a_ka_{N-k-2}
&{\displaystyle\sum_{k=0}^{N-1}}a_ka_{N-k}\cr
\vdots &\vdots     &\vdots &\vdots &\vdots\cr
0      &a_0a_{N-1} &\cdots
&{\displaystyle\sum_{k=0}^{N-3}}a_ka_{2N-k-4}
&{\displaystyle\sum_{k=0}^{N-1}}a_ka_{2N-k-2}}
\nonumber\\
&+&
\pmatrix{
0      &0      &\cdots &0          &0     \cr
a_0^2  &a_1a_0 &\cdots &a_{N-2}a_0 &a_Na_0\cr
\vdots &\vdots &\cdots &\vdots     &\vdots\cr
{\displaystyle\sum_{k=0}^{N-2}}a_ka_{N-k-2}
&{\displaystyle\sum_{k=1}^{N-1}}a_ka_{N-k-1} &\cdots
&{\displaystyle\sum_{k=N-2}^{2N-4}}a_ka_{2N-k-4}
&{\displaystyle\sum_{k=N}^{2N-2}}a_ka_{2N-k-2}}.
\end{eqnarray}
Taking the scalar product, the first and second terms give
$\partial_z\sigma_{N}{}_{\young{1}}$
and $a_0\sigma_N$, respectively, and the third term vanishes.
Hence we have
\begin{equation}
\sigma_N{}_{\young{2}}+\sigma_N{}_{\young{11}}
=\left(\frac{d^2}{dz^2}+z\right)\sigma_N.
\end{equation}

Next we consider the following equality,
\begin{equation}
\sigma_{N}{}_{\young{2}}-\sigma_{N}{}_{\young{11}}
=\pmatrix{
a_2    &a_3    &\cdots &a_{N+1}\cr
a_3    &a_4    &\cdots &a_{N+2}\cr
\vdots &\vdots &\ddots &\vdots \cr
a_{N+1}&a_{N+2}&\cdots &a_{2N}}
\cdot
\pmatrix{
\Delta_{11} & \Delta_{12} &\cdots &\Delta_{1N}\cr
\Delta_{21} & \Delta_{22} &\cdots &\Delta_{2N}\cr
\vdots      & \vdots      &\ddots &\vdots\cr
\Delta_{N1} & \Delta_{N2} &\cdots &\Delta_{NN}\cr}.
\label{eqq1}
\end{equation}
The first matrix of the right hand side of (\ref{eqq1}) is
rewritten as
\begin{eqnarray}
&{}&
\pmatrix{
\partial_z a_1 &\partial_z a_2    &\cdots &\partial_za_{N}  \cr
\partial_z a_2 &\partial_z a_3    &\cdots &\partial_za_{N+1}\cr
\vdots         &\vdots            &\ddots &\vdots           \cr
\partial_z a_N &\partial_z a_{N+1}&\cdots &\partial_za_{2N-1}}
\nonumber\\
&+&
\pmatrix{
a_0^2 &a_0a_1+a_1a_0 &\cdots &{\displaystyle\sum_{k=0}^{N-1}}a_ka_{N-k-1}\cr
a_0a_1&a_0a_2+a_1a_1 &\cdots &{\displaystyle\sum_{k=0}^{N-1}}a_ka_{N-k}  \cr
\vdots&\vdots        &\vdots &\vdots                                     \cr
a_0a_{N-1}&a_0a_N+a_1a_{N-1} &\cdots
&{\displaystyle\sum_{k=0}^{N-1}}a_ka_{2N-k-2}}
\nonumber\\
&+&
\pmatrix{
0      &0      &\cdots &0     \cr
a_1a_0 &a_2a_0 &\cdots &a_Na_0\cr
\vdots &\vdots &\cdots &\vdots\cr
{\displaystyle\sum_{k=1}^{N-1}}a_ka_{N-k-1}
&{\displaystyle\sum_{k=2}^{N}}a_ka_{N-k} &\cdots
&{\displaystyle\sum_{k=N}^{2N-2}}a_ka_{2N-k-2}}.
\label{eqq2}
\end{eqnarray}
Here, we note that $a_n$'s also satisfy
\begin{equation}
\partial_za_{n+1}=2na_{n-1}\ ,\label{recursion2}
\end{equation}
which is proved by induction from eqs.(\ref{orecursion}) and (\ref{oini}).
The first term of the right hand side of eq.(\ref{eqq2}) is rewritten
by using eq.(\ref{recursion2}) as
\begin{eqnarray}
&{}&
\pmatrix{
0      &2a_0    &\cdots &2(N-1)a_{N-2}\cr
0      &2a_1    &\cdots &2(N-1)a_{N-1}\cr
\vdots &\vdots  &\vdots &\vdots       \cr
0      &2a_{N-1}&\cdots &2(N-1)a_{2N-3}}
\nonumber\\
&+&
\pmatrix{
0             &0             &\cdots &0       \cr
2a_0          &2a_1          &\cdots &2a_{N-1}\cr
\vdots        &\vdots        &\cdots &\vdots  \cr
2(N-1)a_{N-2} &2(N-1)a_{N-1} &\cdots &2(N-1)a_{2N-3}}.
\label{eqq3}
\end{eqnarray}
Applying the scalar product on these terms, we obtain
\begin{equation}
\sigma_N{}_{\young{2}}-\sigma_N{}_{\young{11}}
=(2N-1)z\sigma_N.
\end{equation}
Hence we have proved lemma \ref{le2}.\hfill \lower5pt\hbox{$\Box$}

Continuing the similar argument, we get the following
shift operators.
\begin{lem}\label{le3}
\begin{eqnarray}
&{}&
\sigma_N{}_{\young{3}}+2\sigma_N{}_{\young{12}}+\sigma_N{}_{\young{111}}
=\left(\frac{d^3}{dz^3}+3z\frac{d}{dz}+1\right)\sigma_N,
\label{shift4}\\
&{}&
\sigma_N{}_{\young{3}}-\sigma_N{}_{\young{111}}
=\left( (2N-1)z\frac{d}{dz}+(2N+1)\right)\sigma_N,
\label{shift5}\\
&{}&
\sigma_N{}_{\young{3}}-\sigma_N{}_{\young{12}}+\sigma_N{}_{\young{111}}
=\left(2z\frac{d}{dz}+2(N^2+N-1)\right)\sigma_N.
\label{shift6}
\end{eqnarray}
\end{lem}

Finally, we prove proposition \ref{prop2}. {}From the Pl\"ucker relations, we have
\begin{equation}
\sigma_{N+1}{}_{\young{11}}\sigma_N
-\sigma_{N+1}{}_{\young{1}}\sigma_N{}_{\young{1}}
+\sigma_{N+1}\sigma_N{}_{\young{2}}=0,
\end{equation}
\begin{equation}
\sigma_{N+1}{}_{\young{12}}\sigma_N
-\sigma_{N+1}{}_{\young{2}}\sigma_N{}_{\young{1}}
+\sigma_{N+1}\sigma_N{}_{\young{3}}=0,
\end{equation}
which are essentially the same as the bilinear equations
(\ref{mKP1}) and (\ref{mKP2}) for $\tau_{N,{\rm KP}}$.
By using lemmas \ref{le1}, \ref{le2} and \ref{le3}, we get 
\begin{equation}
D_{z}^2\sigma_{N+1}\cdot\sigma_N=0\ ,
\end{equation}
\begin{equation}
\left(D_{z}^3 + 4zD_z - 4(N+1)\right)\sigma_{N+1}
\cdot\sigma_N=0 \ ,
\end{equation}
which are the desired result. Thus we have proved proposition \ref{prop2}.
\renewcommand{\thesection}{\Roman{section}.}
\section{Concluding Remarks}
In this article, we have presented two types of determinant representations
for the rational solutions of P$_{\rm II}$.
The Devisme polynomial representation follows from
the reduction procedure of modified KdV equation
and the Hankel determinant representation is obtained from
the Toda lattice equation,
namely the B\"acklund transformation of the solution of P$_{\rm II}$.
These determinant structures of the rational solutions of P$_{\rm II}$
exactly reflect the Wronskian structure of the solution of KP
hierarchy and Toda lattice equation. The relationship between
those two representations is not clear yet. At least, it seems that
there is no simple transformation relating the two representations.

It is known that the Airy function type solutions of P$_{\rm II}$ are
expressed as\cite{Okamoto}
\begin{equation}
v=\frac{d}{dz}\log\frac{\rho_{N+1}}{\rho_N}\ ,
\end{equation}
\begin{equation}
\rho_N=\left| \matrix{
Ai & \frac{d}{dz} Ai & \cdots & \frac{d^{N-1}}{dz^{N-1}}Ai\cr
\frac{d}{dz}Ai & \frac{d^2}{dz^2} Ai & \cdots & \frac{d^{N}}{dz^{N}}Ai\cr
\vdots & \vdots & \ddots &\vdots \cr
\frac{d^{N-1}}{dz^{N-1}}Ai & \frac{d^N}{dz^N} Ai & \cdots & \frac{d^{2N-2}}{dz^{2N-2}}Ai\cr
}\right|\ ,\label{airy}
\end{equation}
where $Ai$ is the Airy function satisfying
\begin{equation}
\frac{d^2}{dz^2}Ai=z~Ai\ .
\end{equation}
Then $v$ satisfies P$_{\rm II}$,
\begin{equation}
\frac{d^2v}{dz^2}=2v^3-2zv+(2N+1)\ .
\end{equation}
In \cite{OandN}, it was shown that the $\tau$ function
(\ref{airy}) can be reduced from that of the KP hierarchy (\ref{tau_KP}).

In the theory of KP hierarchy, an important fact is that we can introduce the
$\tau$ function which is expressed in terms of determinant.
Based on this fact, we can identify the solution space, and 
the KP hierarchy is regarded as the dynamical system on the 
infinite dimensional Grassmann manifold.
P$_{\rm II}$ is obtained from the similarity
reduction of the modified KdV equation, but the parameter of the
equation appears as the integration constant, which means that
P$_{\rm II}$ has the information of various boundary conditions
of the modified KdV equation. {}From this observation, it looks that
one cannot expect such beautiful structures in the solution space of
P$_{\rm II}$. Nevertheless, the results in this
article may imply that at least for the special function type solutions and
the rational solutions, such structures
in the solutions of KP hierarchy may survive through the reduction.
It may be an interesting problem to investigate the determinant
structures for other Painlev\'e equations. So far, this is completely
an open problem.

Recently, discrete versions of the Painlev\'e equations have been
proposed through the singularity confinement test\cite{dP}. As for the
solutions, some of them admit discrete or $q$-difference
analog of special function type solutions expressed by
determinants\cite{dP2},\cite{dP3}. Moreover, it was reported that
the discrete Painlev\'e II equation admits rational solutions
with determinant structure\cite{dP2rational}.
We might expect that through such determinant structures of solutions,
similarity reductions\cite{SRG} deriving the discrete Painlev\'e equations
from discrete KP (or Toda) would become more transparent,
as we have seen in the continuous P$_{\rm II}$ case.

\ \par
\noindent
{\bf Acknowledgement}\\
The authors are grateful to Profs. J. Satsuma, B. Grammaticos and  A. Ramani
for discussions and encouragement. One of the authors(K.K.) also thanks
Prof. K. Okamoto for leading attention to refs.\cite{Devisme},\cite{Erd} and
useful comments.
This work has been partially
supported by the Research Promotion Funds of Doshisha University.

\end{document}